# The Wave Vector is Limited from Below

I. A. Stepanov
Department of Engineering Materials, University of Sheffield, Sir Robert Hadfield Building, Mappin Street, Sheffield, S1 3JD, UK. Email: igstepanov@yahoo.com

## Abstract

It is shown that the electron wave vector in a crystal cannot be smaller than a certain value. As a result, in a crystal there are many first Brillouin zones, and not only one as it had been supposed earlier. In crystal there can be many electrons with the same wave vector, and not only two as it had been supposed earlier. These electrons must be at sufficiently large distances from each other. An attempt is made to explain the decrease of electric and thermal conductivity with decreasing crystal size.

The wave vector of an electron in a crystal can be written in the form

$$\mathbf{k} = \frac{m_1}{N_1}\mathbf{b}_1 + \frac{m_2}{N_2}\mathbf{b}_2 + \frac{m_3}{N_3}\mathbf{b}_3 \tag{1}$$

where $\mathbf{b_i}$ are the reciprocal lattice vectors, $N_i$ are the numbers of atoms at the edges of the crystal, and $m_i$ are integers. Apparently, with the increase of $N_i$ the wave vector tends to zero. It can be assumed that it cannot be arbitrarily small:

$$\mathbf{k} = \sum_i \frac{m_i}{N_i}\mathbf{b}_i, \quad N_i \leq N_{i0}, \tag{2}$$

$$\mathbf{k} = \mathbf{k_0} = \sum_i \frac{m_i}{N_{i0}}\mathbf{b}_i, \quad N_i > N_{i0}$$

i.e., the wave vector decreases to a certain bound, and starting from certain values of $N_i$ it remains constant. This means that the wave length in a crystal cannot be arbitrarily big.

In [2, 3] it is shown that crystals consist of cells whose entropies are mutually independent. The cells are macroscopic regions of regular shape whose internal energy is much greater than the energy of their interaction with the other parts of the body. Fluctuations in the cells are independent of each other. It is shown in [2, 3] that the mean quadratic fluctuation of a quantity f depends on the number of particles in the crystal according to the following law:

$$\left\langle (f - \bar{f})^2 \right\rangle = \text{const}, \quad N \geq N_c, \tag{3}$$

$$\left\langle (f - \bar{f})^2 \right\rangle \propto N, \quad N < N_c$$

where $N_c$ is the number of atoms in a cell of independent entropy. That is the mean quadratic fluctuation is proportional to N for $N < N_c$. The sizes of the cells of independent entropy are known for whiskers [2, 3]. For this type of crystals of inorganic substances the diameter of the cell is about $10^{-5}$ m and its length is about $4 \cdot 10^{-5}$ m. Consequently, the volume of the cell is approximately $3.14 \cdot 10^{-13}$ m$^3$, which corresponds to a cube with an edge length of about $6.8 \cdot 10^{-5}$ m.

Earlier the following hypothesis on solid state quanta had been stated [4, 5]: "for each crystal solid there exists a minimum quantity by which the body preserves its own normal properties, i.e., the properties of a big sample".

One can assume that $N_c = N_{10}N_{20}N_{30}$, i.e., the wave vector decreases only in the range of the cell of independent entropy. This leads to the following conclusions:

- In a crystal there is not one first Brillouin zone, but many: one for each cell of independent entropy.
- In a crystal there can be many electrons with the same wave vector, and not only two as is had been thought earlier. These electrons can be found in different cells of independent entropy.

*Certainly, there exist crystals with a size of a few meters and a weight of a few tons. It is unlikely that an electron at one face of such a crystal can "feel" that there is another electron with the same wave vector at the opposite face. Here it is necessary to refine the uncertainty principle. Earlier it was thought that an electron with a given wave vector is "smeared" over the whole crystal. The author doubts this. If we agree with this point of view, then an electron in the ice of the Antarctica should be smeared over the ice of the whole Antarctica. This is unlikely, since in this case it would have such a small density that it could not exist at all. An arbitrary object can exist only until its density is greater than a given value. An electron with a given wave vector will be smeared over a finite volume of ice, the only question is the size of this volume. It is necessary to refine the understanding of infiniteness. There is no infinity, there is only a distance which is thought of as infinitely big. For instance, theoretically the interaction of two hydrogen atoms equals zero at infinity. Actually, this interaction equals zero already at a distance of 1 cm. 1 cm means infinity in this case. It is not surprising at all that an electron with a given wave vector will be "smeared" over the crystal in a volume not greater than 1 cm$^3$. For the microscopic world this is an enormous distance. The author believes that an electron is "smeared" over the volume of a cell of independent entropy. This corresponds to $10^5$ lattice constants.*

- The number of wave vectors in the first Brillouin zone equals the number of elementary cells in the cell of independent entropy, and not in the whole crystal.
- It is possible to calculate the crystal, knowing the cell of independent entropy alone. By calculating the cell of independent entropy, we obtain information about the whole crystal.

Earlier it was noted several times that the physical properties of substances do not change with decreasing sample size until a certain sample size has been reached. Then these properties change with further decrease of the sample size. That is how strength [2, 3], electric conductivity, heat conductivity etc. behave [4, 5]. For example, the strength increases. It is proportional to the mean quadratic fluctuation in the energy of the molecule [2, 3, 6]:

$$\left\langle (\Delta E)^2 \right\rangle^{1/2} = \left( \frac{4\pi^4 N'^2}{5\theta^3 N} T^5 \right)^{1/2}, \tag{4}$$

where θ is the Debye temperature, N' is the quantity of atoms in the fluctuation volume, and N is the quantity of atoms in the crystal, which is proportional to the number of wave vectors in the first Brillouin zone. Consequently, the strength should continuously increase as the size of the crystal decreases. Moreover, it is known that the strength does not change until the size of a few microns is reached, and begins to increase only from this size.

According to (4), this means that with a decrease of the sample size, the number of wave vector values, which is proportional to N, does not change until a certain sample size has been reached. Then the number of wave vector values begins to decrease with a decrease of the sample size. This only means that the wave vector in the first Brillouin zone is bounded from below: the density of states in this zone has the form

$$\rho = \frac{V_c}{8\pi^3} = \text{const}, \quad V \geq V_c, \tag{5}$$

$$\rho = \frac{V}{8\pi^3}, \quad V < V_c$$

where $V_c$ is the volume of a cell of independent entropy. In [2, 3] it is shown that the strength of a sample begins to increase when the size of the sample becomes smaller than a cell:

$$\left\langle (\Delta E)^2 \right\rangle^{1/2} = \left( \frac{4\pi^4 N'^2}{5\theta^3 N_c} T^5 \right)^{1/2} = \text{const}, \quad N \geq N_c \tag{6}$$

$$\left\langle (\Delta E)^2 \right\rangle^{1/2} = \left( \frac{4\pi^4 N'^2}{5\theta^3 N} T^5 \right)^{1/2}, \quad N < N_c$$

The thermal conductivity and the electric conductivity of the crystal decrease starting from a given sample size. The coefficient of thermal conductivity k is proportional to the heat capacity $C_V$ and the density of the one-electron levels at the Fermi surface $g(E_F)$ [1]:

$$k \propto C_V \propto g(E_F) \tag{7}$$

The density of states in metals does not differ very much from its value for a free electron gas, with the exception of transition metals:

$$g(E_F) \propto \frac{3N}{2E_F V} \tag{8}$$

According to (8), the thermal conductivity should remain unchanged with decreasing sample size. In (8) there is a multiplier 1/V because in the derivation of (8) there was a multiplication by 8π/V, i.e., by the volume of the reciprocal space, corresponding to one wave vector. One can write (8) in the form

$$g(E_F) \propto \frac{3N_c}{16\pi^3 E_F} \frac{8\pi^3}{V_c} \tag{9}$$

With the decrease of the sample, g(E_F) remains unchanged. One can assume that just when the size becomes smaller than a cell of independent entropy, the number of wave vectors in the first Brillouin zone will decrease, but the volume of the reciprocal space, corresponding to one vector, will not increase. I.e., the crystal "remembers" the volume corresponding to one vector, and preserves it. The electric conductivity is proportional to the thermal conductivity according to the Wiedemann–Franz law.

# ВОЛНОВОЙ ВЕКТОР ОГРАНИЧЕН СНИЗУ

## И. А. СТЕПАНОВ

*Институт химической физики, Латвийский Университет, Рига, LV-1586, Латвия*

Показано, что волновой вектор электрона в кристалле не может быть меньше определённого числа. В результате этого в кристалле много первых зон Бриллюэна, а не одна, как считалось ранее. В кристалле могут быть много электронов, обладающих данным значением волнового вектора, а не два, как считалось раньше. Эти электроны должны находиться на достаточно большом расстоянии друг от друга. Сделана попытка объяснить уменьшение электропроводности и теплопроводности с уменьшением размера кристалла. Теория подтверждается опытом.

It has been shown that the electron wave vector in crystal can not be smaller than a certain value. As the result in crystal there are many first Brillouin zones but not one, as it has been supposed earlier. In crystal there can be many electrons with certain value of the wave vector, but not two, as it has been supposed earlier. These electrons must be at sufficiently large distance from each other. An attempt has been made to explain the decreasing of electric and thermal conductivity with decreasing of the size of the crystal.

Волновой вектор электрона в кристалле можно записать в следующем виде [1]:

$$\mathbf{k} = (m_1/N_1)\mathbf{b_1} + (m_2/N_2)\mathbf{b_2} + (m_3/N_3)\mathbf{b_3}, \qquad (1)$$

где $\mathbf{b_1}$ - векторы обратной решётки, $N_1$ - число атомов на рёбрах кристалла, $m_1$ - целые числа. Видно, что с увеличением $N_1$ волновой вектор неограниченно убывает. Можно предположить, что он не может быть как угодно малым:

$$\mathbf{k} = \sum_i (m_i/N_i)\mathbf{b}_i; \quad N_i \leq N_{i0},$$
$$\mathbf{k} = \mathbf{k_0} = \sum_i (m_i/N_{i0})\mathbf{b}_i; \quad N_i > N_{i0}, \qquad (2)$$

то есть вектор убывает до определённого предела, а, начиная с определённого значения $N_i$ остаётся постоянным. Это значит, что длина волны в кристалле не может быть как угодно большой.

В [2,3] показано, что кристаллы состоят из клеток, энтропия которых независима друг от друга. Клетки - это макроскопические ячейки правильной формы, флуктуации в них происходят независимо друг от друга. В [2,3] показано, что средняя квадратичная флуктуация какой - либо величины $f$ зависит от числа частиц в кристалле по такому закону:

$$\left\langle (f - \bar{f})^2 \right\rangle = const, \quad N \geq N_c,$$
$$\left\langle (f - \bar{f})^2 \right\rangle > \sim N, \quad N < N_c, \qquad (3)$$

где $N_c$ - число атомов в клетке независимой энтропии. Размеры клеток независимой энтропии известны для нитевидных кристаллов [2, 3]. Для нитевидных кристаллов неорганических веществ диаметр клетки $\approx 10^{-5}$ м, длина $\approx 4 \cdot 10^{-3}$ м. Следовательно, объём ячейки $\approx 3{,}14 \cdot 10^{-13}$ м$^3$, это куб со стороной $\approx 6{,}8 \cdot 10^{-5}$ м.

Ранее уже высказывалась гипотеза о квантах твёрдого вещества [4,5]: "для каждого кристаллического твёрдого тела существует некоторое минимальное количество, при котором это тело ещё сохраняет свои нормальные свойства, то есть свойства массивного образца".

Можно предположить, что $N_c = N_{10} N_{20} N_{30}$, то есть волновой вектор убывает только в пределах клетки независимой энтропии. Это приводит к таким выводам.

- В кристалле не одна первая зона Бриллюэна, а много первых зон Бриллюэна: одна для каждой клетки независимой энтропии.
- В кристалле могут быть много электронов, обладающих данным значением волнового вектора, а не два, как считалось ранее. Эти электроны находятся в разных клетках независимой энтропии.

*Действительно, есть кристаллы размером несколько метров и весом несколько тонн. Маловероятно, что электрон у края такого кристалла "чувствует", что у другого края есть электрон с таким же волновым вектором. Тут надо уточнить соотношение неопределённости. Ранее считалось, что электрон с определённым значением волнового вектора "размазан" по всему кристаллу. Автор сомневается в этом. Согласно этой точке зрения, электрон во льду Антарктиды "размазан" по всему льду Антарктиды. Это маловероятно, так как в таком случае он имел бы такую малую плотность, что не мог бы существовать. Любой объект существует только до тех пор, покуда его плотность не меньше определённой величины. Электрон с определённым значением волнового вектора будет "размазан" по определённому конечному значению объёма льда, вопрос по какому? Надо уточнить понятие бесконечности. Бесконечности нет, есть расстояние, которое считается бесконечно большим. Например, теоретически взаимодействие двух атомов водорода равно нулю на бесконечности. Фактически, взаимодействие равно нулю на расстоянии 1 см. 1 см - это бесконечность для данного случая. Нет ничего удивительного в том, что электрон с определённым значением волнового вектора в кристалле будет "размазан" по объёму 1см$^3$. Для микромира это астрономически большая величина. Автор считает, что электрон будет "размазан" по объёму клетки независимой энтропии. Это $10^5$ постоянных решётки.*

- Количество волновых векторов в первой зоне Бриллюэна равно количеству элементарных ячеек в клетке независимой энтропии, а не во всём кристалле.
- Можно рассчитать кристалл, зная только клетку независимой энтропии. Рассчитав клетку независимой энтропии, получаем информацию о всём кристалле.

Ранее неоднократно замечалось, что физические свойства вещества не изменяются при уменьшении размеров образца, а, начиная с некоторого размера, начинают изменяться. Так ведёт себя прочность [2,3], электропроводность, теплопроводность и т. д. [4,5]. Прочность, например, увеличивается. Прочность пропорциональна средней квадратичной флуктуации энергии молекулы [2,3,6]:

$$\left\langle (\Delta E)^2 \right\rangle^{1/2} = \left( (4\pi^4 N'^2 / 5\theta^3 N) \cdot T^5 \right)^{1/2} , \qquad (4)$$

где $\theta$ - температура Дебая, $N'$ - количество атомов во флуктуационном объёме, а $N$ - количество атомов в кристалле, пропорциональное числу волновых векторов в первой зоне Бриллюэна. Следовательно, прочность должна постоянно возрастать с уменьшением размеров кристалла. Однако известно, что прочность не изменяется до микронных размеров и лишь затем начинает возрастать.

Согласно (4), это значит, что с уменьшением размера образца количество значений волнового вектора, пропорциональное $N$, не изменяется, а, начиная с некоторого размера, убывает. Это говорит только о том, что волновой вектор в первой зоне Бриллюэна ограничен снизу: плотность состояний в этой зоне имеет вид

$$\rho = V_c / 8\pi^3 = const, \ V \geq V_c ,$$
$$\rho = V / 8\pi^3, \ V < V_c , \qquad (5)$$

где $V_c$ - объём клетки независимой энтропии. В [2,3] показано, что прочность образца начинает изменяться, когда размер образца становится меньше клетки:

$$\left\langle (\Delta E)^2 \right\rangle^{1/2} = \left( (4\pi^4 N'^2 / 5\theta^3 N_c) \cdot T^5 \right)^{1/2} = const, \quad N \geq N_c ,$$

$$\left\langle (\Delta E)^2 \right\rangle^{1/2} = \left( (4\pi^4 N'^2 / 5\theta^3 N) \cdot T^5 \right)^{1/2}, \quad N < N_c .$$
(6)

Теплопроводность и электропроводность кристалла уменьшаются, начиная с определённого размера образца. Коэффициент теплопроводности *к* пропорционален теплоёмкости $C_V$ и плотности одноэлектронных уровней у поверхности Ферми $g(E_F)$ [1]:

$$k \sim C_V \sim g(E_F).$$
(7)

Плотность состояний в металлах не слишком сильно отличается от её значений для свободного электронного газа, за исключением переходных металлов:

$$g(E_F) \sim 3N / 2E_F V.$$
(8)

Согласно (8), теплопроводность должна оставаться неизменной с уменьшением размера образца. В (8) есть множитель $1/V$ потому, что при выводе (8) было умножение на $8\pi/V$, то есть на объём в обратном пространстве, соответствующий одному волновому вектору. Можно переписать (8) в виде

$$g(E_F) \sim (3N_c / 16\pi^3 E_F) 8\pi^3 / V_c .$$
(9)

При уменьшении образца $g(E_F)$ остаётся постоянной. Можно предположить, что как только размер станет меньше клетки независимой энтропии, количество волновых векторов в первой зоне Бриллюэна будет уменьшаться, но объём обратного пространства, приходящийся на один вектор, увеличиваться не будет. То есть, кристалл "запоминает" объём, приходящийся на один вектор и сохраняет его. Электропроводность пропорциональна теплопроводности по закону Видемана-Франца.

**Литература**